\documentclass[12pt]{iopart}
\usepackage {amssymb}
\newcommand{\nc}{\newcommand}
\nc{\ba}{\begin{eqnarray}}
\nc{\ea}{\end{eqnarray}}
\nc{\rb}{\bar{\rho}}
\nc{\p}{\phi}
\nc{\la}{\lambda}
\nc{\al}{\alpha^{\prime}}
\nc{\de}{\delta_{H}}
\nc{\be}{\begin{equation}}
\nc{\ee}{\end{equation}}
\nc{\D}{\overline{\mbox{D}}}
\input{epsf.sty}

\begin{document}


\title{Brane Inflation and Cosmic String Tension in Superstring Theory}

\vspace{1cm}

\author{Hassan Firouzjahi \footnote{Electronic mail:
firouzh@lepp.cornell.edu}
and S.-H.~Henry Tye \footnote{Electronic mail:
tye@lepp.cornell.edu}}
\vspace{0.5cm}
\address
{Laboratory for Elementary Particle Physics, Cornell University, 
Ithaca, NY 14853}

\date{January 7, 2005}

\begin{abstract}
In a simple reanalysis of the KKLMMT scenario, we argue that the slow roll 
condition in the D3-$\D$3-brane inflationary scenario in superstring 
theory requires no more than a moderate tuning. The cosmic string 
tension is very sensitive to the conformal coupling: with less fine-tuning,
the cosmic string tension (as well as the ratio of tensor to scalar perturbation mode) 
increases rapidly and can easily saturate the present observational bound. 
In a multi-throat brane inflationary scenario, this feature substantially improves 
the chance of detecting and measuring the properties of the cosmic strings as a 
window to the superstring theory and our pre-inflationary universe. 
  
\vspace{0.3cm}

Keywords : Cosmic strings, Inflation, Superstring Theory, Cosmology
\end{abstract}

\section{Introduction}

Cosmic strings were originally proposed as an alternative to inflation, providing the seeds 
for structure formation in our universe \cite{Vilenkin}. The cosmic microwave background data from WMAP and earlier experiments \cite{cobe,wmap} strongly 
supports the inflationary universe scenario \cite{guth} to be the origin of the early universe, 
and rules out the cosmic string scenario for structure formation. 
However, the origin of the inflaton and its corresponding inflaton potential are not well understood.

Branes play a key role in superstring theory \cite{Polchinski}, and the brane world is a natural 
realization of our universe in string theory. In this scenario, standard model particles are open string modes confined to the branes (which have 3 large spatial dimensions) while the gravitons are closed string modes in the bulk.
Brane inflation \cite{dvali-tye}, where non-BPS branes move towards each other in the brane 
world, is a generic inflationary scenario in superstring theory.
In this scenario, inflation ends when the branes collide and heat the universe, initiating 
the hot big bang. Cosmic strings (but not domain walls or monopoles) are
copiously produced during the brane collision \cite{{Jones:2002cv},{Sarangi:2002yt}}. 
Properties of superstring theory cosmic strings are well studied 
\cite{Jones:2003da,Pogosian:2003mz,Kachru:2003sx,Copeland:2003bj,
Leblond:2004uc,Jackson:2004zg,Kibble:2004hq}.
Their contribution to density perturbations is small 
compared to that coming from the inflaton. Their properties are compatible with all observations
today, but are likely to be tested in the near future. The properties of the cosmic strings depend 
on the type of branes involved in the scenario. So detecting and measuring the properties of the 
cosmic strings provides a window to the superstring theory and our pre-inflationary universe. 

The simplest brane inflationary scenario is realized  by 
considering the motion of a D3-brane moving towards an $\D$3-brane 
in a compact manifold \cite{collection}. This scenario can be realized by 
introducing an extra D3-${\D}$3-brane pair in a KKLT-like vacuum \cite{Kachru:2003aw},
where all moduli are dynamically stabilized. The $\D$3-brane is sitting 
at the bottom of a throat (the A throat) as the mobile D3-brane is moving towards it. 
This is the KKLMMT scenario \cite{Kachru:2003sx}.
For an inflationary scenario to work, the production of defects other than cosmic 
strings must be suppressed by many orders of magnitude. The D3-${\D}$3 scenario achieves 
this property automatically. A simple generalization of the above 
model, namely, the multi-throat scenario \cite{Burgess:2004kv,Chen:2004gc},
can be easily achieved by putting the standard 
model branes in another throat (the S throat), where the gauge hierarchy problem may be 
solved with enough warping \cite{Randall:1999ee}, while the inflationary 
constraints require a more modest warping for the A throat. 
Recent arguments \cite{Barnaby:2004gg} suggests that 
heating up the standard model branes to start the hot big bang may not be an issue.
We believe this multi-throat scenario is quite generic and realistic.
In addition, it also allows one to ensure the (initial) stability of cosmic 
strings \cite{Copeland:2003bj,Leblond:2004uc} so that they can evolve into a scaling network \cite{Wyman} that may be detected. 
Here we shall consider this scenario in some detail.

Consider the following generic potential in the A throat in the multi-throat scenario,
\be
\label{infpot}
V= V_K + V_A + V_{D \bar D} = \frac{1}{2}\beta H^2 \phi^2  
+ 2T_3h_A^4(1-\frac{1}{N_A}\frac{\phi_A^4}{ \phi^4}) + ...
\ee
where the first term $V_K$ receives contributions from the K\"{a}hler potential and various 
interactions in the superpotential \cite{Kachru:2003sx} as well as possible D-terms
\cite{Burgess:2003ic}. $H$ is the Hubble constant and the inflaton $\phi$ is the position of 
the D3-brane, so this interaction term behaves like a conformal coupling. In general, $V_K$  
depends on where the D3-brane is sitting so $\beta$ is actually a function of $\phi$. 
However, we expect $\beta$ to stay more or less constant in each throat, so we use that 
approximation here. 
$\beta$ receives many contributions and is non-trivial to 
determine \cite{Shandera:2004zy,Berg:2004sj, Bobkov}. 
Generically $\beta \sim 1$; for slow roll, $|\beta |$ must be small; this is a fine-tuning.
The second term $V_A$ is the effective cosmological 
constant coming from the presence of the $\D$3-brane (with tension $T_3$)
sitting at the bottom of the A throat. $h_A \ll 1$ is the corresponding warp factor. 
This term drives inflation.
The last term $V_{D \bar D}$ is the Coulombic-like attractive potential between the D3-brane 
and the $\D$3-brane. 

Generically we expect $\beta \sim 1$ and the slow roll condition for enough 
e-folds of inflation is not satisfied. Unless the D3-brane sits in a trench with a 
shift symmetry \cite{Firouzjahi:2003zy,Shandera:2004zy}, the model requires a fine-tuning 
on $\beta$. In the KKLMMT scenario, $|\beta | \lesssim 1/100$ so that $V_{D \bar D}$ dominates 
over $V_K$. This is the 2 orders of magnitude fine-tuning alluded to in Ref\cite{Kachru:2003sx}.
Here we would like to ask exactly how small $\beta$ must be to satisfy the slow roll condition
and other constraints in inflation.
We find that $\beta \simeq 1/6$ is possible, in which case $V_K$ (actually all 3 terms in $V$)
plays an important role in the inflationary scenario. This means only a moderate tuning of $\beta$
is needed. Note that $\beta$ depends on where the throat is sitting in the bulk.
If there are many throats, it is likely that one of the throats will have a small enough $\beta$, so
a throat with $\beta \lesssim 1/6$ may not require any tuning. In this case, we may start with a number of 
extra pairs of D3-$\D$3- branes. The D3-brane around the throat with the smallest $\beta$ is 
likely to move the slowest and ends up as the brane responsible for inflation.

One may understand the $\beta$ relaxation simply by looking at the slow roll parameter 
$\eta = \beta/3 - C/\p^6$, where $C$ is a constant obtainable from Eq(\ref{infpot}). 
Suppose that $\eta$ is positive at the 
beginning of the inflationary epoch. As $\p$ decreases, the cancellation between the 2 terms 
guarantees that $\eta$ will be very small for some range of $\p$. 
For relatively large $\beta$, $\eta$ would be positive at $N_e \simeq 55$
e-folds before the end of inflation, so the density perturbation power spectrum index 
$n_s \simeq 1+2\eta >1$,
that is, the scale-invariant spectrum is blue-tilted. In comparison, the spectrum is red-tilted
when $\beta=0$.
 
What may be most interesting in the above $\beta$ relaxation is that the cosmic string tension 
$\mu$ grows rapidly with increasing $\beta$.  
For (p,q) strings \cite{Copeland:2003bj}, we find (with $G$ as the Newton's constant and $g_s$ the string coupling)
\be
\label{Gmu1}
\hspace{2cm}G\mu_{p,q} = 4 \times 10^{-10}\sqrt{p^2g_s + q^2/g_s}\, f({\beta})
\ee
where $f(\beta)$ is a function of $\beta$ and $N_e$, where $N_e \simeq 55$ is the number 
of e-folds before the end of inflation.
For $\beta=0$, $f(\beta=0)=1$, so $G \mu$ reduces to that obtained for the KKLMMT scenario
\cite{Kachru:2003sx,Copeland:2003bj}, where either the F-string ((p,q)$=$(1,0)) or the D-string ((p,q)$=$(0,1)) has 
$G \mu \ge 4 \times 10^{-10}$. For large $\beta N_e$, $f(\beta)$ grows like 
$(\beta N_e)^{5/4}\exp (\beta N_e/2)$, so $G\mu$ grows rapidly even with moderate values of $\beta$. The cosmic D-string tension  and the power spectrum index $n_s$ as functions of $\beta$ are 
shown in Figure 1.  

As $\beta$ becomes negative, the $\beta$ term will tend to move the D3-brane out of the throat.
To make sure that the attractive D3-$\D$3 force is strong enough to pull the D3-brane
towards the $\D$3-brane at the botttom of the throat (so inflation can end), 
$\beta$ cannot be too negative. It turns out that this condition requires any negative $\beta$ to be exponentially close to zero. 

The WMAP analysis of inflationary parameters \cite{Peiris:2003ff} gives $n_s \le 1.28$, which means
 $\beta\le 0.4$, which is too weak to be useful. 
 The observational bound on the ratio $r$ of the tensor to the scalar mode 
 perturbation, $r\le 0.89$, translates to $\beta < 0.22$, which is not stringent at all. 
 Here $r$ as a function of $\beta$ is shown in Figure 2.
Note that $r$ increases exponentially as $\beta$ increases.  
The observational bound on the running of $n_s$, namely, $d\,n_s/d\, \ln k$, however, is 
not very conclusive. $d\,n_s/d\, \ln k$ as a function of $\beta$ is also shown in Figure 2.
If one uses only the WMAP data, the bound is $d\,n_s/d\, \ln k \le 0.03$, which gives no bound on
$\beta$.
If one uses the WMAP + 2dFGRS data, the bound $d\,n_s/d\, \ln k \le 0.01$ gives
the bound $\beta \lesssim 1/5$. If we take seriously 
\footnote[1] {In contrast to the WMAP data, the cosmological application of the Lyman $\alpha$ 
data involves some non-trivial astrophysics and may incur unknown uncertainties. 
It is important to understand better the Lyman $\alpha$ clouds because of its 
implication to inflation data analysis.}
the bound from WMAP+2dFGRS+Lyman $\alpha$ 
on $d\,n_s/d\, \ln k$, then we find $\beta \lesssim 1/22$.
For $\beta \simeq 1/22$, $G\mu \simeq 10^{-8}/\sqrt g_s$. 
However, in the more recent analysis of WMAP data combined with SDSS Lyman $\alpha$ and other data \cite{Seljak}, the 
inflationary parameters are determined. Depending on how they fit the data, they find $n_s$ 
goes from $0.98 \pm 0.02$ to $1.02 \pm 0.033$. If the bound on $n_s$ is taken to be 2 $\sigma$ 
above the larger central value, we find $n_s \le 1.086$, or around $\beta \le 1/7$.  
In this paper, we shall use this bound on $\beta$.
It is clear that more data and additional analysis 
will be important. If the data/analysis holds up, extensions of this scenario should also be 
explored thoroughly. Generically, any model extension will in general relax the bounds on the parameters. 

We see that, in this  D3-${\D}$3-brane inflationary scenario,
the cosmic string tension has its value in the range 
\be
\label{Gmub}
\hspace{3.2cm} 4 \times 10^{-10}  \lesssim G \mu \lesssim 6 \times 10^{-7}
\ee
where a fine-tuning is needed to reach the lower values and  
the upper value is the present observational bound \cite{Pogosian:2003mz,smoot}
coming from WMAP data \footnote[2]
{Bound on the gravitational wave background from pulsar timing also puts a bound on $G \mu$.
However, they are not as stringent \cite{Kaspi}. Ref\cite{Lommen} apparently gives a
stronger bound than that quoted here. However, that bound assumes a gravitational wave 
background at one frequency, which is suitable for massive binary black holes but not 
for cosmic strings, which have a very wide frequency band. We thank E. Flanagan for 
pointing out this issue.}. 
This observational upper bound gives $\beta \lesssim 1/7$.

It is interesting that the constraints on $\beta$ ($\lesssim 1/6$) from model building (enough e-folds of inflation etc.) is quite close to that from the data. 
For practical purposes, we shall use
\ba
\hspace{3.2cm} 0 \lesssim \beta \lesssim 1/7
\ea
We note that slight variations of the above scenario can relax the bounds further.
We summarize the various bounds in Table 1.

With larger $G \mu$, more ways to detect cosmic strings are opened up.
The signature of the cosmic strings coming from the superstring theory typically has
a non-trivial spectrum in the tension \cite{Jones:2003da,Copeland:2003bj}. 
Larger $G \mu$ drastically improves the chances of detecting and measuring the 
tension spectrum and so the testing of superstring theory.

Larger $G\mu$ also corresponds to a blue tilt of the scale invariant
power spectrum of the density perturbation. As $\beta$ increases (say $\beta \ge 0.02$)
the power spectrum index $n_s $ becomes blue-tilted. 
For small $G \mu$, $\beta$ is negligibly small so the power spectrum 
is red-tilted, as given by the KKLMMT scenario. 
We note that the range $G \mu$ in this D3-${\D}$3-brane inflationary scenario is 
comparable to that given in Ref\cite{Sarangi:2002yt} for a variety of brane inflationary scenarios.
There, the larger $G \mu$ values correspond to the branes-at-a-small-angle 
scenario \cite{Garcia-Bellido:2001ky}. In that scenario, the power spectrum is red-tilted, so a measurement of $n_s$ can distinguish between these 2 scenarios. 
Measuring the cosmic string tension spectrum also will be important to distinguish them.  
In general, measuring $n_s$, $r$, $d\,n_s/d\, ln k$ together with the $G \mu$ tension spectrum and the corresponding number densities will go a long way to pinpoint the particular brane inflationary scenario.
We note that the predictions of the above inflationary quantities and the cosmic string tension 
are insensitive to the warp factor $h_A$ and the background charge $N_A$. For example 
$10^{-3} \lesssim h_A \lesssim 10^{-1}$  and $N_A \lesssim 10^4$  are compatible with the 
the inflationary scenario and the above predictions. Note that $N_A \gg 1$ is assumed 
for the validity of the supergravity approximation used. For our purpose, $N_A \ge 100$.

\vspace{1.3cm}
\centerline{\vspace{-0.3cm}\hspace{-12cm}\small{Log($G\mu$)}}\centerline{\hspace{3cm}\vspace{0cm}
\small{$n_s-1$}}
\centerline{\epsfxsize=6.5cm\epsfbox{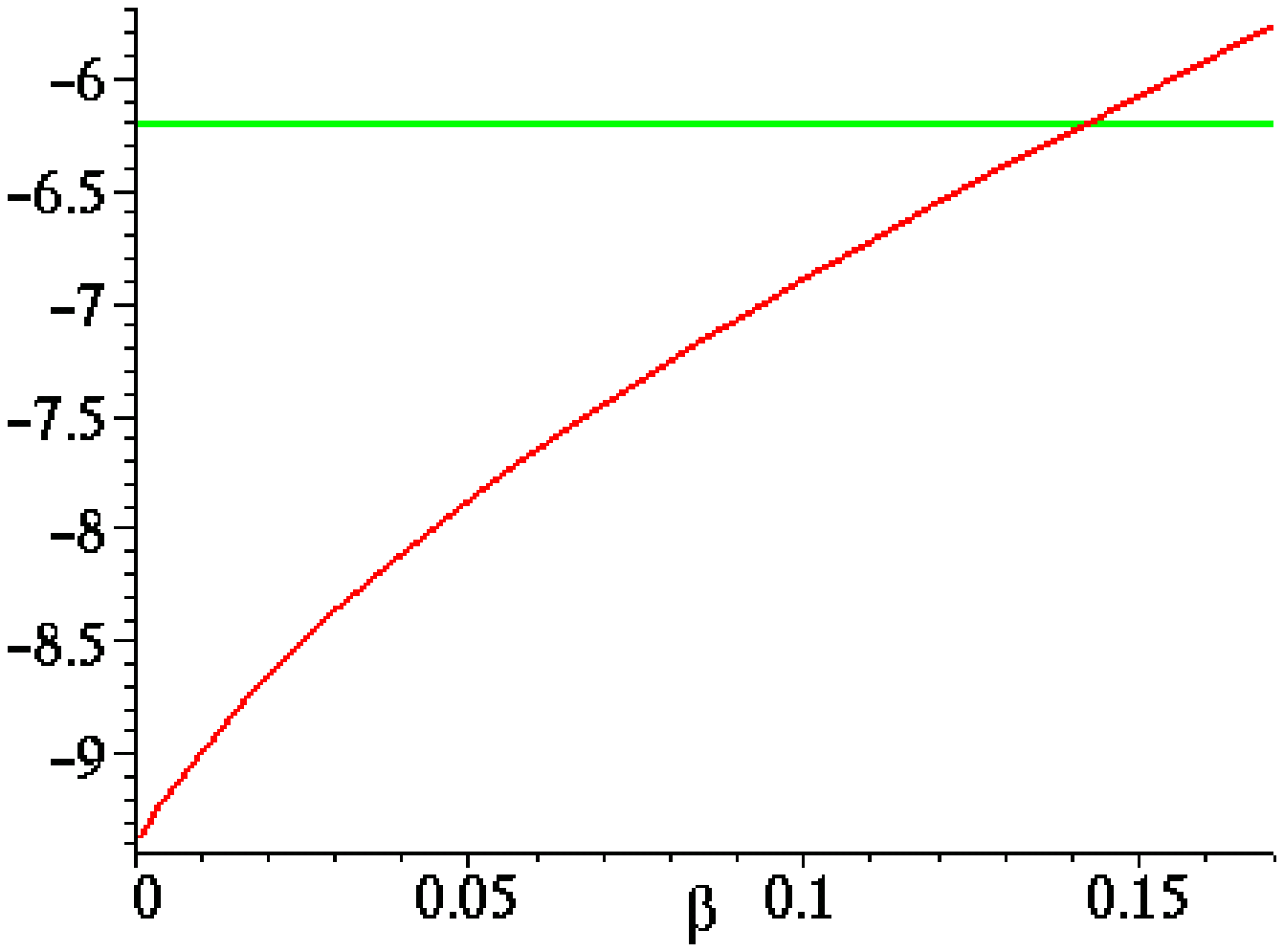}\hspace{1cm} \epsfxsize=6.5cm\epsfbox{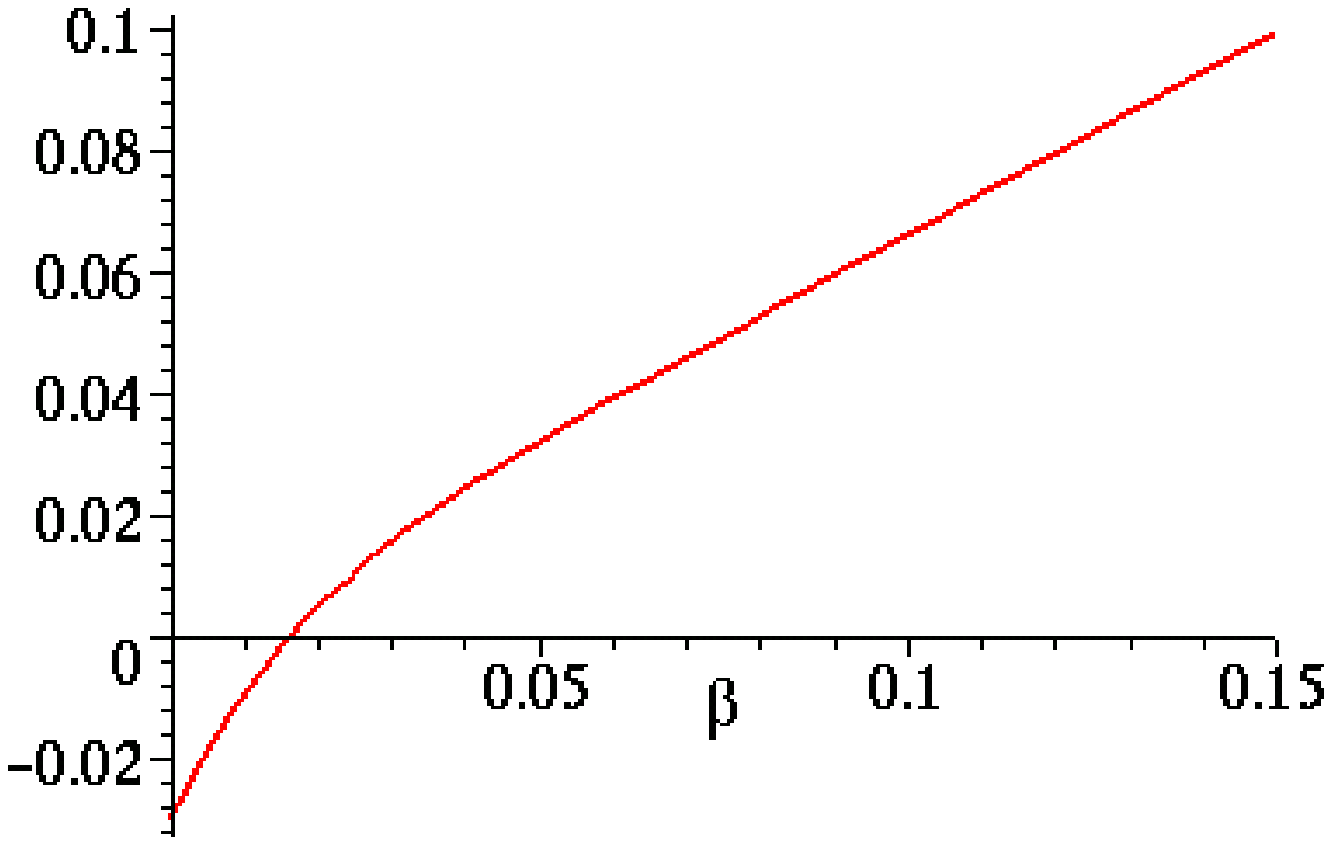}}
\vspace{0.2in}
\noindent {{\bf FIG 1}. The cosmic string tension $G\mu$ and the density perturbation power 
spectrum index $n_s$ are shown as functions of $\beta$ in the KKLMMT-like 
brane inflationary scenario. Here $g_s=1$.
The horizontal line at -6.2 in the Log($G\mu$) graph indicates the present observational bound
on the cosmic string tension, which corresponds to
$\beta \lesssim 1/7$. The observational constraint on $n_s \le 1.086$ also gives $\beta \lesssim 1/7$. 

\label{1}

\vspace{1.5cm}

\centerline{\vspace{-0.5cm}\hspace{-13cm}\small{Log $r$}}\centerline{\hspace{3cm}
\small{$d\,n_s / d\, \ln k$}}
\centerline{\epsfxsize=6.5cm\epsfbox{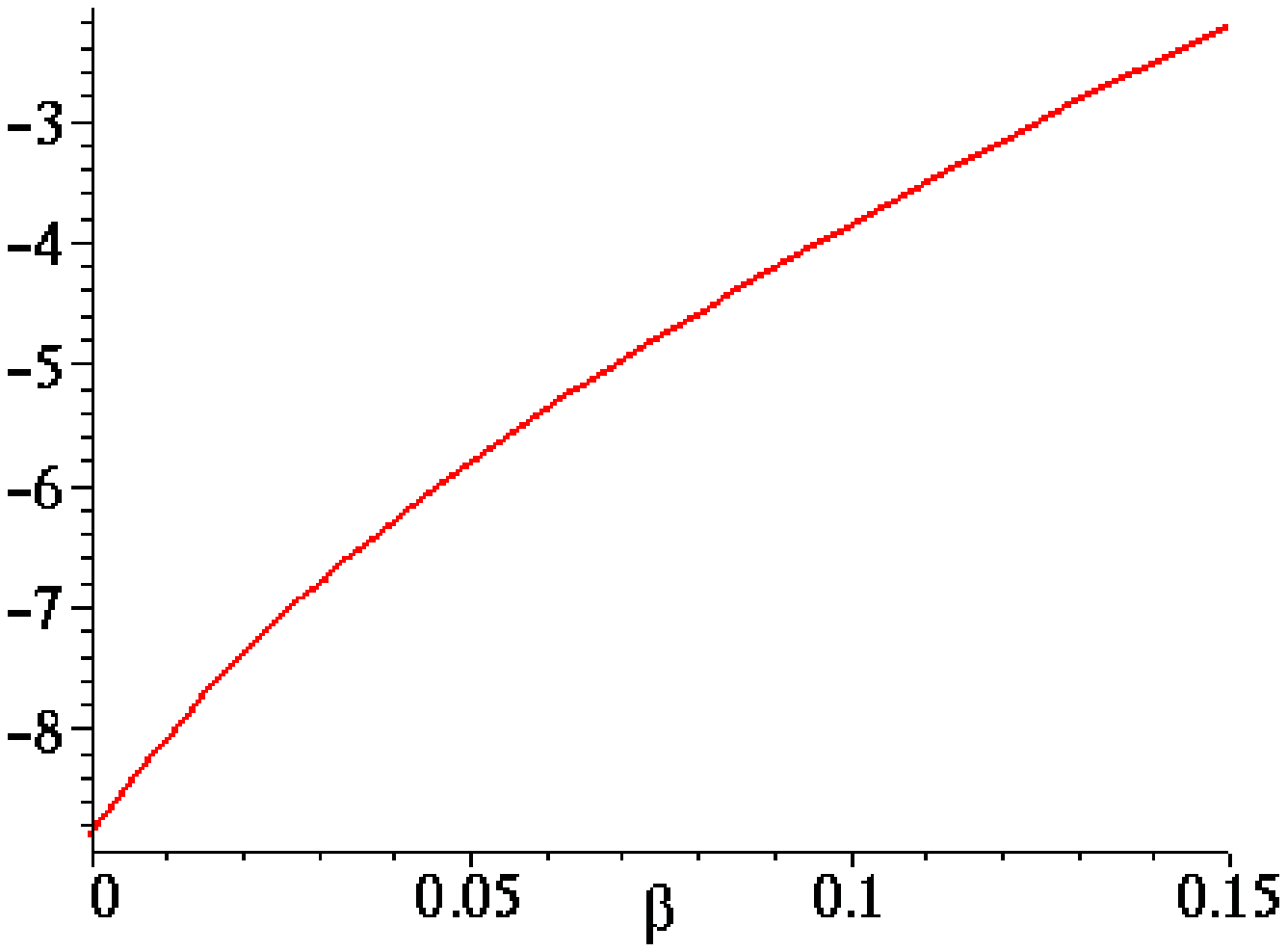}\hspace{1cm} \epsfxsize=6.5cm\epsfbox{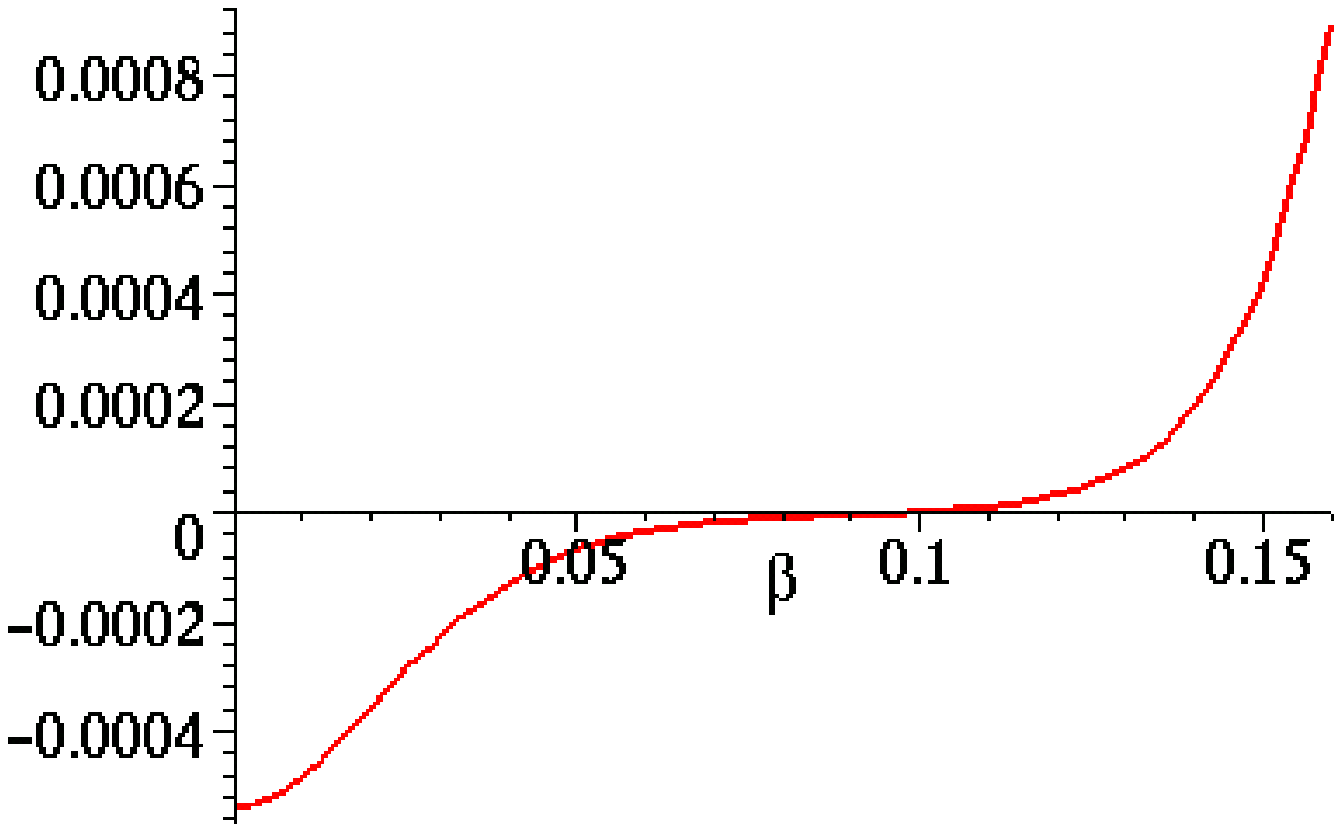}}
\vspace{0.15in}
\noindent {{\bf FIG 2}.The ratio of the tensor to scalar perturbations, $r$, and
the running of the density perturbation power spectrum index, $d\,n_s/d\, \ln k$, are shown as functions of $\beta$.}
\label{2}

\vspace{2cm}

\centerline{
\begin{tabular}{|l|l|r|}  \hline
model building& $\beta \lesssim $ 1/6\\ \hline\hline
\hspace{1.1cm}$n_s$ & $\beta \lesssim 1/7$ \\ \hline
\hspace{1.1cm}$r$ & $\beta \lesssim 1/5$ \\ \hline
\hspace{1cm}$G\,\mu$ & $\beta \lesssim 1/7$ \\ \hline
\end{tabular}}
\vspace{0.2in}
\noindent {{\bf Table 1.} Various bounds on $\beta$ coming from observational 
data \cite{Peiris:2003ff,Pogosian:2003mz,smoot,Seljak}. 
The more recent analysis \cite{Seljak} using all available data gives $n_s \le 1.086$ which implies $\beta \lesssim 1/7$.}
\vspace{0.8cm}

Fast roll in the multi-throat scenario was proposed as an interesting alternative to 
slow roll \cite{Chen:2004gc,Alishahiha:2004eh}. However, this fast roll scenario requires
a fine-tuning (e.g., the background charge $N \sim 10^{14}$), so the above 
slow roll scenario seems most natural. For larger values of $\beta$ at the A throat, we 
also consider the possibility that inflation takes place while the D3-brane is moving 
slowly out of another throat. In this case, $\beta$ for this 3rd throat must be moderately
small and negative.

\section{The Setup}

The realistic setup is a Type IIB orientifold (or F theory) compactified on a Calabi-Yau
3-fold with fluxes \cite{Giddings:2001yu,Kachru:2003aw}, where all moduli are stabilized.
Inside the bulk of the Calabi-Yau manifold, there are local regions, or throats, with 
warped geometry \cite{Klebanov:2000hb}. 
The metric in any throat has the approximate $AdS_5 \times X_5$ form, where $X_5$ is 
some orbifold of $S_5$ and the $AdS_5$ metric in Poincare coordinates takes the form
\ba
\label{metric}
\hspace{2cm}ds^2=h(r)^2\left(-dt^2 +a(t)^2 d\vec{x}^2 \right)+ h(r)^{-2} dr^2 
\ea
with the warp factor
\ba
\hspace{3cm}h(r)=\frac{r}{R}= e^{-2 \pi K/3Mg_s}\ ,
\ea
where $K$ and $M$ are the background NS-NS and RR fluxes respectively, and $R$ 
represents the curvature radius of the AdS throat and is given by \cite{Klebanov:2000hb}
\ba
\hspace{3cm}R^4=\frac{27}{4}\pi g_s N {\al}^2\, . 
\ea
Here $N=KM$ is equal to the number of the background D3 charge or, equivalently, 
the products of NS-NS and RR flux as constructed in \cite{Giddings:2001yu}. $N$ is taken to 
be relatively large.
There can be a number of such throats in the compact region. 
As long as the tadpole cancellation imposed 
on the charge conservation is satisfied, there is no restriction on the 
number of throats.  Following the convention of Ref\cite{Chen:2004gc} we consider a 
scenario with at least 2 throats : the A throat, where the ${\D}$3-brane is 
located, and the S throat, where the standard model branes are located. As a D3-brane
moves towards the ${\D}$3-brane at the bottom of the A throat, inflation takes place.
The position of the D3-brane is ${\bf r}$ in the 6-dimensional compact space. Around a throat,
we choose the coordinate with respect to the bottom of the throat. This allows us to consider
only $r=|{\bf r}|$. Note that $\phi= \sqrt{T_3} r$ is the inflaton, where
$T_3$ is the D3-brane tension. 
 
In another scenario, the D3-brane can simply be a mobile
brane in the bulk, or it can be produced in another throat (the B throat) via the flux-brane 
annihilation mechanism \cite{DeWolfe:2004qx}.  For the D3-brane to 
move out of the B throat, $\beta$ there must be negative. Each of these throats has 
its own warping which we denote by $h_A$, $h_B$ and $h_S$ respectively.

In slow roll inflation, the D3-brane is moving slowly so that the square root term in 
DBI action of the D-brane world volume action can be expanded
up to quadratic order. One can easily see that the Chern-Simons and
the gravitational part of the DBI action for the D3-brane cancel each other and we are left 
with a canonically normalized kinetic term for $\phi$. 

There are a number of contributions which affect
the motion of the D3-brane either in the bulk or in a throat. An important
one comes from the K\"{a}hler modulus stabilization \cite{Kachru:2003sx}. Other 
contributions come from  a variety of sources \cite{Shandera:2004zy}.
The leading term of these contributions on the effective potential has the form
\ba
\label{Vcorrection}
\hspace{3cm}V_K= \frac{1}{2} \beta \, H^2 \p^2 
\ea
where $\beta(\rho, \phi)$ is a function of the K\"{a}hler modulus $\rho$ and depends on 
where the brane is sitting in the compactified manifold. Around any particular throat,
the $\phi$ dependence is small and may be neglected, so we can treat $\beta$ as a constant.


We assume that there is an $\overline{\mbox{D}}$3-brane located in the A throat. Its tension
provides the vacuum energy for inflation, given by
\ba
\label{V_A}
\hspace{2cm}V_A=2 h_A ^4\, T_3 \, = \frac{2 h_A ^4}{(2\pi)^3 g_s {\al} ^2}.
\ea
The factor of 2 is due to equal contributions of Chern-Simons and the gravitational part of the action 
for $\overline{\mbox{D}}$3-brane. The factor $h_A ^4$ represents the warping effect. 
The resulting inflaton potential for this inflaton field  
comes from Eqs (\ref{Vcorrection}) and (\ref{V_A}) and from the attractive potential $V_{D\bar{D}}$ between the D3 and the $\overline{\mbox{D}}$3-branes,
\ba
\label{V}
\hspace{2cm}V&=&\frac{1}{2} \beta \, H^2 \p^2  + 2 h_A ^4\, T_3 + V_{D\bar{D}}   \nonumber\\
\hspace{2cm}&=& \frac{1}{2} \beta \, H^2 \p^2 +  \frac{64\pi^2 \p_A ^4}{27\,N}\left(1-\frac{1}{N} \frac{{\p_A}^4}{\p^4}\right)\, .
\ea
where $\p_A=\sqrt{T_3}r_0$ is the location of the $\overline{\mbox{D}}$3-brane
at the bottom of the A throat.
In the KKLMMT scenario,
$\beta$ is fine-tuned to 1 part in 100 so that its effect is negligible. In this case,
the doubly-warped interaction $V_{D\bar{D}}$ is easily 
compatible with the slow roll conditions. 

At the end of inflation, the D3-brane collides with the ${\D}$3-brane at the bottom of 
the A throat. Part of the energy produced will be transferred to the S throat to start the hot big bang,
while D1-branes (i.e., D-strings) and fundamental closed strings are also produced. The 
quantity of interest is the D-string tension $G\mu$, where $\mu$ is the effective 
tension measured from the four dimensional effective action point of view. It is related to 
the intrinsic tension $T_1$ of the D1-brane via
\ba
\label{Gmu}
\hspace{2cm}G \mu=G T_1 h_A ^2 \, 
=\sqrt{\frac{1}{32\pi\, g_s}}\left(\frac{T_3}{M_P^4}\, h_A^4 \right)^{\frac{1}{2}}
\ea
where $G^{-1} = 8 \pi M_P^2$ and $g_s$ is the string coupling.

\section{The Scenario}

Here we investigate the range of $\beta$ allowed by the inflationary constraints, in 
particular the slow roll condition in the A throat. Within the allowed range of $\beta$, we 
evaluate the cosmic string tension and various inflationary parameters. 
The KKLMMT scenario is reproduced in the 
$\beta \rightarrow 0$ limit.

To satisfy the slow roll condition, the slow roll parameter $\eta$ must be small enough :
\ba
\label{eta}
\hspace{2cm}\eta=M_P^2 \frac{V''}{V} =\frac{\beta}{3}-\frac{20}{N_A}\frac{{M_P}^2\p_A^4}{\p^6}
\ea
It is easy to check that the other slow roll parameter $\epsilon$ is always small, so we shall 
ignore it in the discussion on slow roll.
Inflation ends when $\eta$ ceases to be small. So we can 
determine the final value of inflaton field, $\p_f$ when $\eta \sim -1$, which gives
\ba
\label{phi_f}
\hspace{2cm}\p_f^6=\frac{1}{(1+\beta/3)}\left(\frac{20}{N_A}\,M_P^2\p_A^4\right)
\ea 
The number of e-folds is given by
\ba
\label{efoldKKLMMT}
\hspace{2cm}N_e = \frac{1}{M_p^2}\int \frac{V d\p}{V'}.
\ea
so the value of $\p$, namely $\p_i$, at $N_e$ before the end of inflation is given by 
\ba
\label{efold}
\hspace{2cm}e^{2\beta N_e}=\frac{\beta N_A\p_i^6+12M_P^2\p_A^4}{\beta N_A\p_f^6+12M_P^2\p_A^4}
\ea
Combining Eqs(\ref{phi_f}) and (\ref{efold}), we find
\ba
\label{phi_i}
\hspace{2cm}\p_i^6&=& \frac{24N_e}{N_A}M_P^2\p_A^4 \, \Omega(\beta) \nonumber\\
\hspace{2cm}\Omega(\beta) &\equiv& \frac{(1+2\beta)e^{2\beta N_e}-(1+\beta/3)}
{2 \beta\,(N_e+5/6) (1+\beta/3)} \nonumber\\
\hspace{2cm}&\simeq & \frac{e^{2\beta N_e}-1}{2 \beta N_e}\, .
\ea
where $\Omega(\beta)$ is normalized such that $\Omega=1$ as $\beta \rightarrow 0$.
The last formula is a good approximation for small $\beta$.
The density perturbation $\de = 1.9 \times 10^{-5}$ measured by cosmic microwave background radiation \cite{cobe} is given by
\ba
\label{COBE}
\hspace{2cm}\de&\equiv & \frac{1}{\sqrt{75}\pi}\frac{1}{M_P^3}\frac{V^{\frac{3}{2}}}{V'}  \nonumber\\
\hspace{2cm} &=& \left(\frac{2^{11}}{3\times 5^6 \times \pi^4}\right)^{\frac{1}{6}}\,
N_e^{\frac{5}{6}}\left(\frac{T_3}{M_P^4}\, h_A^4 \right)^{\frac{1}{3}}\,f(\beta)^{-\frac{2}{3}}
\ea
where $f(\beta)$ is given by
\ba
\label{f}
f(\beta)&=&\left[\frac{2\beta (N_e + 5/6)}{(1+2\beta)e^{2\beta N_e}-(1+\beta/3)}\right]^{5/4}
\frac{(1+2\beta)^{3/2}}{(1+\beta/3)^{1/4}}\, e^{3\beta N_e} \nonumber\\
& \simeq &  \left[\frac{2\beta N_e}{e^{2\beta N_e}-1}\right]^{5/4} e^{3\beta N_e}
\ea
where the last formula is a good approximation for small $\beta$.
In the limit $\beta \rightarrow 0$, $f(\beta) \rightarrow 1$ and our results reduce to those
of KKLMMT. 

One can easily show that
\ba
\label{phi_iphi_R}
\hspace{2cm}\frac{\p_i}{\p_R}=\left(\frac{32 \pi^2}{27N_A}\right)^{\frac{1}{4}}\, 
\left(\frac{T_3}{M_P^4}\right)^{-\frac{1}{4}}\, \left(\frac{\p_i}{M_P}\right) \, 
\ea
where
\ba
\label{phi_iM_P}
\hspace{2cm}\frac{\p_i}{M_P}=\left( \frac{3^{13}\times 5^{6}}{2^{15}}\right)^{\frac{1}{12}}\,
\Omega(\beta)^{\frac{1}{6}}f(\beta)^{\frac{1}{3}}\, N_e^{-\frac{1}{4}}\de^{\frac{1}{2}}\, .
\ea
and $\p_R \equiv\sqrt{T_3} R$. 
Now we can write
\ba
\hspace{2cm}\eta &=& \frac{\beta}{3} - \frac{5}{6N_e} \frac{1}{\Omega(\beta)} \nonumber \\
\hspace{2cm}\epsilon &=& \frac{M_P^2}{2}\left(\frac{V'}{V}\right)^2
= \frac{1}{18}\left(\frac{\phi_i}{M_P}\right)^{2}
\left(\beta + \frac{1}{2N_e \Omega}\right)^2 \nonumber \\
\hspace{2cm}\zeta &=& M_P^4 \frac{V' V'''}{V^2} =\frac{5}{3N_e \Omega}\left(\beta + \frac{1}{2N_e \Omega}\right)
\ea
where is $\p_i/M_P$ is given in Eq(\ref{phi_iM_P}). 
Note that, given COBE normalization $\de =1.9 \times 10^{-5}$ at $N_e \simeq 55$,  $\eta$, $\epsilon$ and $\zeta$ are functions of $\beta$ only.
In general, $(\phi_i/M_P)^2 \beta^2 << 1$, so $\epsilon$ may be neglected. Let us focus on this case, where the slow roll condition is guided by the slow roll parameter $\eta$. 
Due to the opposite sign of the 2 contributions in $\eta$, we see that the presence of the 
$\beta>0$ term actually improves the slow roll condition. This property allows a small but not 
necessarily fine-tuned $\beta$. The $\eta$ at $N_e$ before the end of inflation measures the deviation 
from the scale invariant power spectrum. The power spectrum index $n_s$ at $N_e \simeq 55$
is given by
\ba
\hspace{2cm}n_s -1\simeq  2\eta -6 \epsilon  \simeq \frac{2\beta}{3} - \frac{5}{3N_e} \frac{1}{\Omega(\beta)}
\ea 
and is plotted as a function of $\beta$ in Figure 1.
For the sake of completeness, both the running of $n_s$,
\ba
\hspace{2cm}\frac{dn_s}{d \ln k} = 16 \epsilon \eta -24 \epsilon^2 -2 \zeta 
\ea
and the  ratio of the tensor to the scalar mode perturbation, $r=12.4 \epsilon$, are plotted as 
functions of $\beta$ in Figure 2. 
Although the ratio $r$ is small, its value $r(\beta \simeq 0.1)\simeq 10^{-4}$ is larger 
by about 5 orders of magnitude than the value obtained in KKLMMT limit, where $r \sim 10^{-9}$. 
This provides another good test of the scenario.

The cosmic string tension is given by
\ba
\hspace{2cm}G\mu &=& \left(\frac{3\times 5^6\times \pi^2}{2^{21}}\right)^{1/4}
g_s^{-1/2}\, {\de}^{3/2}\, N_e^{-5/4} \, f(\beta)\, \nonumber\\
\hspace{2cm}&=& G \mu_{0}\, f(\beta)
\ea
where $\mu_0$ is the cosmic D-string tension at $\beta=0$ (the KKLMMT scenario) and has the value \cite{Kachru:2003sx,Copeland:2003bj} \footnote[1]{after correcting a typo in (C.9) in Ref \cite{Kachru:2003sx}},
\ba
\hspace{2cm}G\mu_{0} \simeq \frac{4}{\sqrt{g_s}} \times 10^{-10}
\ea
Both $G \mu$ and $n_s$ as a function of $\beta$ are shown in Fig. 1.
The range of values for $G\mu$ is given in Eq(\ref{Gmub}), where the upper bound comes 
from WMAP data  \cite{Pogosian:2003mz,smoot}. Actually, this bound is obtained 
for abelian Higgs model vortices. For (p,q) strings, the analysis is clearly more complicated.
It is reasonable to assume that the (p,q) string network also evolves to a scaling solution. It is also reasonable to assume that the high tension (i.e., high (p,q)) string densities are suppressed, so that
the range (\ref{Gmub}) may be approximately applied to $G\mu$ for the F- and/or the D-string.

Next, we need to find the bounds on $\beta$ such that the constraints from the total number of 
e-folds (which must be larger than $N_e$, though in general, we expect it to be much larger than $N_e$), the COBE normalization and the consistency of the inflationary model building
in the A throat are satisfied. It turns out that the last condition puts the most stringent upper bound on
$\beta$.
We demand that
\ba
\hspace{3cm}\p_A< \p_f < \p_i < \p_R 
\ea
This condition is to make sure that the last 55 e-folds take place while the D3-brane is moving inside the A throat. The condition (\ref{phi_iphi_R}) can easily be satisfied. One may want to impose the more 
common condition $\p_i/M_P<1$. Using Eq(\ref{phi_iM_P}), one finds for $N_e=55$, 
$\beta \lesssim 1/6$.

As mentioned above, COBE normalization and $n_s$ put weaker bounds on $\beta$. 
To see that, we note from Eq(\ref{COBE}) that $\de$ is given
in terms of three undetermined parameters: $T_3/M_P^4$, $h_A$ and $\beta$.
One can get a crude upper bound for $\beta$ as follows. 
One expects that $T_3/M_P^4 <1$ and $h_A<1$. Consider the extreme case when
$h_A^4\, T_3/M_P^4 =1$. For $N_e \simeq 55$ and $\de=1.9\times 10^{-5}$, 
we find $\beta < 0.46$. 
The constraints of WMAP data on inflationary parameters are presented in Ref\cite{Peiris:2003ff}.
The observational bound on $n_s$ 
for $\eta > 0$ and $ n_s >1$ is $1.00 \le n_s \le$ 1.28 \cite{Peiris:2003ff}. 
This in turn implies that $\beta < 0.4$. 
The observational bound on $r$ 
is $r \le 0.89$ \cite{Peiris:2003ff}. This gives the bound $\beta < 0.22$. 
The observational bound on $d\,n_s/d\, \ln k$, however, is more restrictive. 
If we take the bound from WMAP+2dFGRS+Lyman $\alpha$ on $d\,n_s/d\, \ln k$ seriously, we 
find that $\beta \lesssim 1/22$ (see Figure 2). For $\beta \simeq 1/22$, 
$G\mu \simeq 10^{-8}/\sqrt{g_s}$. However, this bound will be relaxed if the number of parameters
to be fitted is increased.
Also, if one uses only the WMAP data on $d\,n_s/d\, \ln k$, the bound is not restrictive at all.
Using the WMAP+2dFGRS data, the bound on $\beta$ is $\beta\lesssim 1/5$, which is still less 
stringent than the other bounds. 
In the recent analysis of WMAP data combined with SDSS Lyman $\alpha$ and other 
data \cite{Seljak}, the inflationary parameters as well as other parameters are determined. 
Depending on how the data are fitted, $n_s$ ranges from $0.98 \pm 0.02$ to 
$1.020 \pm 0.033 + 0.066 -0.061$, where the last 2 numbers are 2 $\sigma$ uncertainties
(note 2 $\sigma$ is not twice 1 $\sigma$). The lower value of $n_s$ imposes no constraint on 
the model. If the upper bound on $n_s$ is taken to be 2 $\sigma$ 
above the larger central value, we find $n_s \le 1.086$, or around $\beta \le 1/7$. The 
bound on $\beta$ from the bounds on $d\,n_s/d\, \ln k$  and $r$ are not as restrictive.
Here, we shall use $\beta \lesssim 1/7$.
However, one may be concerned about the astrophysical uncertainties involved in applying 
the Lyman $\alpha$ data here. This issue should wait for more data as well as analysis. 

Having obtained the upper bound on $\beta$, we now determine the lower bound on it. The strongest
lower bound comes from the requirement that the D3-brane moving in the outskirts of the throat
eventually moves toward $\overline{\mbox{D}}$3-brane so inflation can end. 
Since the D3-brane is moving very slowly this can
be achieved by an attractive force, which means $V'|_{\p=\p_R} >0$. We have
\ba
\hspace{1.5cm}V'|_{\p=\p_R}=\frac{\p_R V_A}{3M_P^2}\left(\beta +16\pi \sqrt{\frac{2}{3}} N_A^{-\frac{3}{2}}
h_A^4 (\frac{T_3}{M_P^4})^{-\frac{1}{2}} \right)\, .
\ea
The combination $N_A^{-\frac{3}{2}}h_A^4$, independent of the details of model, is 
very small. This indicates that an attractive force at the beginning of the throat 
for a negative $\beta$ is not possible unless $\beta$ is exponentially close to zero. 
This reduces to the KKLMMT scenario.
To conclude, we find that $0 \lesssim  \beta \lesssim 1/7$.

Let us summarize the situation with the parameters. 
With $N_e \simeq 55$, there are at least 5 parameters, namely,
$h_A$, $g_s$, $N_A$, $\beta$ and $T_3/M_P^4$. The physically measurable parameters are
$G \mu (\beta, g_s), \de(\beta, h_A^4T_3/M_P^4)$, $n_s(\beta)$, $r(\beta, \de)$ and 
$d\,n_s/d\, \ln k (\beta, \de)$. 
It is interesting to realize that the measurable cosmological quantities are insensitive to $h_A$
and $N_A$. For example, using Eq(\ref{COBE}), it is easy to show that $h_A\,(T_3/M_P^4)^{1/4}$
is approximately equals to $10^{-4}$ and $10^{-3}$ at $\beta=0$ and $\beta=0.1$, respectively.
This indicates that a range of $10^{-3}\le h_A \le 10^{-1}$ is possible with the appropriate 
value of $T_3/M_P^4$. 
An upper bound on $N_A$ can be obtained by imposing that inflation ends 
before the distance between D3 and $\D$3 reaches $\sqrt{\alpha'}$. At this separation,
tachyon appears and inflation must end as in hybrid inflation. However, inflation can end earlier.
Using metric (\ref{metric}) to 
calculate the physical distance between D3 and $\D$3, this condition translates to
\ba
\hspace{3cm}\frac{\p_f}{\p_A} \ge \exp(\sqrt{\alpha'}/R) \, .
\ea
Using Esq (\ref{phi_f}) and (\ref{COBE}) in above expression, we find
\ba
\label{N_Abound}
N_A\, \exp \left( (4/27\pi g_sN_A )^{1/4}\right) \le \left(\frac{2^{29}}{5^{2}
\times3^{7}}\right)^{1/6}\,\frac{ N_e^{5/6}}
{\left(1+\beta /3 \right)^{2/3}
f(\beta)^{2/3}\, \de} \nonumber
\ea
For example this gives $N_A \lesssim 10^4$ at $\beta =1/6$ and  $\lesssim 10^7$ at $\beta=0$.
Recall that $N_A \gg 1$ in the supergravity approximation we are using. Typically $N_A \ge 100$ should be sufficient. 

\section{Slow Roll in B throat}

In this section we consider the case where $\beta$ is not small enough in the A throat. We imagine
that the slow roll inflation take places in the B throat which has sufficiently small $\beta$.
We begin with the inflaton potential (\ref{infpot}). $V_A$ is as given in Eq(\ref{V_A}), coming 
from the $\D$3-brane locating at the bottom of the A throat. $V_K$ has the same form as in 
Eq(\ref{Vcorrection}) but with $\beta \rightarrow \beta_{B}$. The interaction term, $ V_{D\bar{D}}$, 
has the combination of warp factors $h_A$ and $h_B$ with the Columbic form $1/(\p-d)^4$ where
now $\p$ is measured with respect to the bottom of the B-throat and $d$ is the distance between 
D3-brane and the $\D$3-brane. The exact form of $ V_{D\bar{D}}$
is a little complicated, but we do not need it in our analysis. This is justified by noting that 
if the A throat and the B throat are not overlapping, $d > R_A + R_B $, one can assume that 
$V_K$ dominates over $ V_{D\bar{D}}$ while the D3-brane is moving out of the B throat. 
When it exits the B throat, $\p \sim \sqrt{T_3} R_B$, $ V_{D\bar{D}}$ starts to become 
comparable to $V_K$ and we take that as the end of B throat inflation. 
For the D3-brane to move out of the B throat, it is necessary that $\beta _B <0$, 
as in Ref\cite{Chen:2004gc}.

The slow roll parameter $\eta \simeq \beta/3$, so to satisfy the slow roll condition, one needs $|\beta|\le 1/20$.
The number of e-folding is given by
\ba
\label{Befol}
\hspace{3cm}\p_i=\sqrt{T_3} R_B\, \exp(-|\beta|\, N_e/3)\, .
\ea
As explained above, it is assumed that inflation ends when the D3-brane reaches the 
top of the B throat, so in Eq (\ref{Befol}) we set $\p_f=\sqrt{T_3} R_B$. Also note that the exponential 
dependence here is weak because $\beta N_e/3 \sim 1$.

The COBE normalization and the cosmic string tension, respectively, are
\ba
\hspace{1.5cm}\de=\left(\frac{2^7}{3\pi^2}\right)^{\frac{1}{4}}\, N_B^{-\frac{1}{4}}\, 
|\beta|^{-1}\, e^{|\beta| N_e/3}\,  h_A^2 \left(\frac{T_3}{M_P^4}\right)^{\frac{1}{4}}
\ea
\ba
\label{Gmu3}
\hspace{1.5cm}G\mu=\left(\frac{3}{2^{17}}\right)^{\frac{1}{4}}\, N_B^{\frac{1}{4}}
\,|\beta|\,e^{-|\beta| N_e/3}
\left(\frac{T_3}{M_P^4}\right)^{\frac{1}{4}} \de 
\ea

One can see that with sufficient warping the observational value for $\de$ is easily
obtained. On the other hand, the degeneracy of the combination $h_A^4\, T_3/M_P^4$ is now
broken in the $G\mu$ expression in the B throat. So far in all calculations in the A throat, $h_A^4$
and $T_3/M_P^4$, although both independent parameters of the model, appeared jointly.
In the B throat inflation, however, the location of the cosmic string and the period of the inflation
are in two different throats which in turn results in the separation of $h_A^4$ 
from $T_3/M_P^4$ in the
$G\mu$ expression. With Eq(\ref{Gmu3}), it is easy to obtain $G\mu$ ranging from
$10^{-10}$ to $10^{-7}$, depending on the value of $T_3/M_P^4$. For example, taking
$T_3/M_P^4 \sim 10^{-7}$, $\beta\sim 1/20$, $N_B \sim 200$, one finds 
$G\mu \sim  1.8\times 10^{-8}$. Decreasing $T_3/M_P^4$ to $10^{-9}$ reduces $G\mu$ to $10^{-9}$.

Due to the attractive NS-NS plus RR forces, $\D$3-branes in the bulk tend to move towards 
the bottom of throats. With enough $\D$3-branes falling into the B throat, brane-flux annihilation takes place \cite{DeWolfe:2004qx}. Recall that $K$ is the NS-NS background flux and $M$ is the RR background flux of the throat. 
For $p \lesssim M$, $p$ $\D$3-branes can cluster to form a maximal size fuzzy NS-5-brane
which then rolls down, leading to the brane-flux annihilation, i.e, the disappearance of the $p$ 
$\D$3-branes. This results in $K \rightarrow K-1$ and the production of $M-p$ D3-branes.  
So it is quite generic to expect a number of D3-branes being produced in the B 
throat \cite{Chen:2004gc}. Inflation takes place as 
these D3-branes slowly move out of the B throat. One may visualize a spray of D3-branes leaving 
the B throat. They may enter different throats in the compact region. Some may enter the A throat while some may enter the S throat. With $k$ numbers of D3-branes colliding with $\bar k$ numbers of $\D$3-branes in the S throat,  the resulting
annihilation will heat up the remaining ${\bar k} - k$ standard model $\D$3-branes. This way,
heating of the standard model branes in the S throat happens more or less simultaneously 
as the annihilation of branes in the A throat. This mechanism may help to solve the reheating problem.

\section{Fast Roll Inflation}

Two different kinds of inflationary scenarios in string theory are studied in the literature, namely,
the conventional ``slow roll'' inflation as in the KKLMMT model and the ``relativistic'' fast roll
inflation in Ref \cite{Silverstein:2003hf,Alishahiha:2004eh} and \cite{Chen:2004gc}. 
Unlike slow roll inflation, one must retain the DBI action for the kinetic term in fast roll inflation. 
It seems that the fast roll inflation 
suffers a serious problem regarding the COBE normalization and the 
non-Gaussianity constraints from the WMAP data. Here we shall comment on
Chen's multi-throat case briefly.

We begin with the DBI action coupled to the gravity
\ba
\label{S}
   S=\frac{1}{2}\int dx^4\, {M_P}^2 \sqrt{-g} \left(R -2V \right)+ S_{DBI} ,
\ea
where 
\ba
\label{DBI}
S_{DBI}=\int dx^4 \sqrt{-g} f(\phi)^{-1}\left(\sqrt{1+f(\phi)
g^{\mu\nu}\partial_{\mu}\p \partial_{\nu}\p} -1 \right)
\ea
where as in \cite{Alishahiha:2004eh}
\ba
\hspace{3cm}f(\p)=\frac{\la}{\p^4} \ .
\ea
Here, $\p=\sqrt{T_3}r$ and $\la \equiv T_3 R^4=27N/32\pi^2$. 

$V$ is now given by $V= V_A +V_K$.
Here we consider that the relativistic D3-brane is moving out of the B throat, which means 
$\beta \sim -1$ \cite{Chen:2004gc}. The positive $\beta$ scenario was studied in 
Ref\cite{Alishahiha:2004eh}.

As discussed in appendix {\bf{B}} of Ref \cite{Alishahiha:2004eh}, the condition that $V$ dominates over
the kinetc energy requires 
\ba
\label{Vdominate}
\hspace{3cm}\frac{V^{3/2}}{V'M_P}\sqrt{\frac{3f}{g_s}} \gg1
\ea
while the condition for relativistic motion ($\gamma \gg 1$) is
\ba
\label{gamma}
\hspace{3cm}\frac{g_s}{3V}V'^2 f M_P^2 \gg 1 \, .
\ea
The number of e-folds is given by 
\ba
\label{efoldChen}
N_e=\int \frac{d\p}{M_P}\sqrt{\frac{fV}{3g_s}}
\sim \sqrt{\frac{2\,\lambda }{3g_s}}\left(\frac{T_3}{M_P^4}\, h_A^4 \right)^{\frac{1}{2}}\left(\frac{M_P}{\p_i}\right)
\ea
The COBE normalization
is given by
\ba
\label{powerspectrum}
\de=\frac{1}{15\pi}\sqrt{\frac{f}{g_s}}\frac{V}{M_P^2}=
\frac{2}{15\pi}\sqrt{\frac{\lambda}{g_s}}\left(\frac{T_3}{M_P^4}\, h_A^4 \right)\left(\frac{M_P}{\p_i}\right)^2
\ea
Eqs(\ref{efoldChen}) and (\ref{powerspectrum}) can be combined to yield
\ba
\label{tuning1}
\hspace{3cm}\frac{\lambda}{g_s}=\frac{N_e^4}{25\pi \, \de ^2}
\ea
Taking $\de ^2 \sim 10^{-10}$ and $N_e=55$, we find
$\lambda/g_s \sim 10^{14}$, or the background D3-brane charge to have an extremely large value :
\ba
\label{tuning2}
\hspace{3cm} N \sim 10^{14}
\ea
which implies a fine-tuning. 
A similar result was also obtained for the case in \cite{Alishahiha:2004eh}; they suggest that 
a non-trivial orbifold can reduce this number to a more reasonable value. It would be interesting 
to see how such a large effective background charge can be achieved in a realistic model. 
It is also possible that the value of $N$ may be substantially reduced by a more careful analysis 
of the brane collision in the infrared end of the throat\footnote[1]{Chen, private communication}. 
This interesting possibility needs further investigation. In any case, such scenarios have large 
non-Gaussianity effects, thus offering a good signature for detection.

The cosmic string tension, using Eqs (\ref{Gmu}) and (\ref{efoldChen}), is
\ba
\label{tensionfastrool}
G\mu=\frac{N_e}{8}\sqrt{\frac{3}{\lambda\, \pi}\,}\left(\frac{\p_i}{M_P}\right)
\sim 7\,\left(\frac{\p_i}{M_P}\right)\times 10^{-7}
\ea

Depending on the value of $\p_i/M_P$, $G\mu$ as big as $10^{-7}$ is possible, while 
$G\mu$ as small as in KKLMMT corresponds to $\p_i/M_P \sim 10^{-3}$, which requires some tuning.

\section{Conclusion}

One may consider a multi-throat brane inflationary scenario
where the brane annihilation takes place in a (the A) throat other than the throat where
the standard model branes are sitting (the S throat). This multi-throat scenario has the advantage
that the warped geometry \cite{Randall:1999ee} of the standard model throat allows 
us to solve the hierarchy problem while the warping of the A throat allows us to fit the 
inflationary data separately. To summarize, we study a simple generalization of the 
original KKLMMT scenario by allowing a small but not necessarily very small conformal coupling. 
More specifically, we consider the following cases :

(1) The original KKLMMT scenario is reviewed, where $|\beta| \ll 1/10$, so that the 
$\beta$ term in the inflaton potential may be neglected. This requires a fine-tuning.
As we relax this fine-tuning to $\beta \lesssim 1/6$, slow roll inflation still works fine, while 
the cosmic string 
tension increases rapidly. In this very simple generalization of the KKLMMT scenario, the
cosmic string tension can easily saturate the observational bound. As the fine-tuning 
is relaxed, the density perturbation power spectrum moves from red-shifted to blue-shifted,
and $r$ increases exponentially.
We also note that the predictions of inflationary quantities and the cosmic string tension 
are insensitive to the warping $h_A$. 

(2) If $\beta$ of the A throat is not small, inflation may take place in another throat where
$\beta$ is small and negative. This can be justified by the brane-flux annihilation picture of 
Ref \cite{DeWolfe:2004qx}. In this scenario, the D3-brane moves out of that (B) throat 
and then drops into the A throat. Most(if not all) inflation takes place when the brane moves 
out of the B throat. In a more general scenario, we expect a number of D3-branes moving out of the B 
throat, which may end up in different throats. It is also possible that inflation happens when the 
$D$3-brane is moving in the bulk towards the A throat.

(3) Suppose that $|\beta| \sim 1$. The positive $\beta$ case is studied in Ref
\cite{Alishahiha:2004eh}
while the negative $\beta$ case is studied in Ref\cite{Chen:2004gc}. In both cases, 
there is no slow roll (in fact, the brane is moving very relativistically) and it is possible
to achieve enough inflation. On the other hand, fitting the density perturbation data
requires $N \sim 10^{14}$, a fine-tuning, although this fine-tuning may be substantially
ameliorated with non-trivial orbifolding. 

Since slow roll and fast roll inflations are both possible, a general analysis of an arbitrary roll will 
be very informative. In the end, we expect the inflationary constraints on $\beta$ 
and the value of the cosmic string tension to be rather relaxed. 

\ack{We thank Xingan Chen, Dave Chernoff, Eanna Flanagan, Andrei Linde, Liam McAllister, 
Joe Polchinski, Sash Sarangi, Sarah Shandera, Eva Silverstein,
Ira Wasserman and Mark Wyman for discussions. This work is supported by the National Science Foundation under Grant No. PHY-009831.}\\


\section*{References}
\begin{thebibliography}{}

\bibitem{Vilenkin} A. Vilenkin and E.P.S.~Shellard,
\underline{Cosmic strings and other topologiocal defects}, 
Cambridge University Press, 2000.

\bibitem{cobe}
G.~F.~Smoot {\it et al.},
``Structure in the COBE DMR first year maps,''
Astrophys.\ J.\  {\bf 396}, L1 (1992); \\
C.~L.~Bennett {\it et al.},
``4-Year COBE DMR Cosmic Microwave Background Observations: Maps and Basic Results,''
Astrophys.\ J.\  {\bf 464}, L1 (1996)
[arXiv:astro-ph/9601067].
 
\bibitem{wmap}
C.~L.~Bennett {\it et al.},
``First Year Wilkinson Microwave Anisotropy Probe (WMAP) Observations:
Preliminary Maps and Basic Results,''
Astrophys.\ J.\ Suppl.\  {\bf 148}, 1 (2003),
astro-ph/0302207; \\ 
D.~N.~Spergel {\it et al.}  [WMAP Collaboration],
``First Year Wilkinson Microwave Anisotropy Probe (WMAP) Observations:
Determination of Cosmological Parameters,''
Astrophys.\ J.\ Suppl.\  {\bf 148}, 175 (2003),
astro-ph/0302209.
                         

\bibitem{guth}
A.~H.~Guth,
``The Inflationary Universe: A Possible Solution To The Horizon And Flatness
Problems,''
Phys.\ Rev.\ D {\bf 23}, 347 (1981); \\
A.~D.~Linde,
``A New Inflationary Universe Scenario: A Possible Solution Of The Horizon,
Flatness, Homogeneity, Isotropy And Primordial Monopole Problems,''
Phys.\ Lett.\ B {\bf 108}, 389 (1982); \\
A.~Albrecht and P.~J.~Steinhardt,
``Cosmology For Grand Unified Theories With Radiatively Induced Symmetry
Breaking,''
Phys.\ Rev.\ Lett.\  {\bf 48}, 1220 (1982).

\bibitem{Polchinski} J.~Polchinski, \underline{String Theory},
        Cambridge University Press, 1998.      

\bibitem{dvali-tye} G.~Dvali and S.-H.H.~Tye,
"Brane Inflation",
Phys. Lett. {\bf B450} (1999) 72, hep-ph/9812483.

\bibitem{Jones:2002cv}
N.~Jones, H.~Stoica and S.-H.~H.~Tye,
``Brane interaction as the origin of inflation,''
JHEP {\bf 0207}, 051 (2002), hep-th/0203163.

\bibitem{Sarangi:2002yt}
S.~Sarangi and S.-H.~H.~Tye,
``Cosmic string production towards the end of brane inflation,''
Phys.\ Lett.\ B {\bf 536}, 185 (2002), hep-th/0204074.

\bibitem{Jones:2003da}
N.~T.~Jones, H.~Stoica and S.-H.~H.~Tye,
'' The production, spectrum and evolution of cosmic strings 
in brane inflation'',
Phys.\ Lett.\ B {\bf 563}, 6 (2003), hep-th/0303269.

\bibitem{Pogosian:2003mz}
L.~Pogosian, S.-H.~H. Tye, I.~Wasserman, and M.~Wyman, 
``Observational constraints on cosmic string production during brane inflation'',  
{\em Phys.  Rev.} D {\bf 68} (2003) 023506,
hep-th/0304188.

\bibitem{Kachru:2003sx}
S.~Kachru, R.~Kallosh, A.~Linde, J.~Maldacena, L.~McAllister
and S.~P.~Trivedi, 
'' Towards inflation in string theory'',
JCAP {\bf 0310} (2003) 013, hep-th/0308055.

\bibitem{Copeland:2003bj}
E.~J.~Copeland, R.~C.~Myers and J.~Polchinski,
``Cosmic F- and D-strings,''
JHEP {\bf 0406}, 013 (2004), hep-th/0312067.

\bibitem{Leblond:2004uc}
L.~Leblond and S.-H.~H.~Tye,
``Stability of D1-strings inside a D3-brane,''
JHEP {\bf 0403}, 055 (2004), hep-th/0402072.

\bibitem{Jackson:2004zg}
M.~G.~Jackson, N.~T.~Jones and J.~Polchinski,
``Collisions of cosmic F- and D-strings,''
hep-th/0405229.

\bibitem{Kibble:2004hq}
T.~W.~B.~Kibble,
``Cosmic strings reborn?,''
astro-ph/0410073; \\
T.~Damour and A.~Vilenkin,
``Gravitational radiation from cosmic (super)strings: Bursts, stochastic
background, and observational windows,''
hep-th/0410222; \\
J.~Polchinski,
``Introduction to cosmic F- and D-strings,''
hep-th/0412244.

\bibitem{collection}
C.~P.~Burgess, M.~Majumdar, D.~Nolte, F.~Quevedo, G.~Rajesh 
and R.~J.~Zhang, JHEP {\bf 07} (2001) 047, hep-th/0105204; \\
G.~R.~Dvali, Q.~Shafi and S.~Solganik,
``D-brane inflation,''
hep-th/0105203; \\      
S.~Buchan, B.~Shlaer, H.~Stoica and S.-H.~H.~Tye,
``Inter-brane interactions in compact spaces and brane inflation,''
JCAP {\bf 0402}, 013 (2004), hep-th/0311207.

\bibitem{Kachru:2003aw}
S.~Kachru, R.~Kallosh, A.~Linde and S.~P.~Trivedi,
'' De Sitter vacua in string theory'',
Phys.\ Rev.\ D {\bf 68}, 046005 (2003),
hep-th/0301240.

\bibitem{Burgess:2004kv}
C.~P.~Burgess, J.~M.~Cline, H.~Stoica and F.~Quevedo,
``Inflation in realistic D-brane models,''
JHEP {\bf 0409}, 033 (2004), hep-th/0403119; \\
N.~Iizuka and S.~P.~Trivedi,
'' An inflationary model in string theory'',
hep-th/0403203.

\bibitem{Chen:2004gc}
X.~Chen,
``Multi-throat brane inflation,''
hep-th/0408084.

\bibitem{Randall:1999ee}
L.~Randall and R.~Sundrum,
``A large mass hierarchy from a small extra dimension,''
Phys.\ Rev.\ Lett.\  {\bf 83}, 3370 (1999), hep-ph/9905221.

\bibitem{Barnaby:2004gg}
N.~Barnaby, C.~P.~Burgess and J.~M.~Cline,
``Warped reheating in brane-antibrane inflation,''
hep-th/0412040.

\bibitem{Wyman} 
A.~Albrecht and N.~Turok,
``Evolution Of Cosmic Strings,''
Phys.\ Rev.\ Lett.\  {\bf 54}, 1868 (1985); \\
D.~P.~Bennett and F.~R.~Bouchet,
``Evidence For A Scaling Solution In Cosmic String Evolution,''
Phys.\ Rev.\ Lett.\  {\bf 60}, 257 (1988); \\
B.~Allen and E.~P.~S.~Shellard,
``Cosmic String Evolution: A Numerical Simulation,''
Phys.\ Rev.\ Lett.\  {\bf 64}, 119 (1990).

\bibitem{Burgess:2003ic}
C.~P.~Burgess, R.~Kallosh and F.~Quevedo,
``de Sitter string vacua from supersymmetric D-terms,''
JHEP {\bf 0310}, 056 (2003), hep-th/0309187.

\bibitem{Shandera:2004zy}
S.~E.~Shandera,
``Slow roll in brane inflation,''
hep-th/0412077.

\bibitem{Berg:2004sj}
M.~Berg, M.~Haack and B.~Kors,
``On the moduli dependence of nonperturbative superpotentials in brane inflation,''
hep-th/0409282.


\bibitem{Bobkov}
K.~Becker, M.~Becker, M.~Haack and J.~Louis,
``Supersymmetry breaking and alpha'-corrections to flux induced potentials,''
  JHEP {\bf 0206}, 060 (2002), hep-th/0204254;
  K.~Bobkov,
  ``Volume stabilization via alpha' corrections in type IIB theory with
  fluxes,'' hep-th/0412239.


\bibitem{Firouzjahi:2003zy}
H.~Firouzjahi and S.-H.~H.~Tye,
``Closer towards inflation in string theory,''
Phys.\ Lett.\ B {\bf 584}, 147 (2004), hep-th/0312020.

\bibitem{Peiris:2003ff}
H.~V.~Peiris {\it et al.},
``First year Wilkinson Microwave Anisotropy Probe (WMAP) observations:
Implications for inflation,''
Astrophys.\ J.\ Suppl.\  {\bf 148}, 213 (2003), astro-ph/0302225.

\bibitem{Seljak}
 U. Seljak, A. Makarov, P. McDonald, S. Anderson, N. Bahcall, J. Brinkmann, S. Burles, R. Cen, 
 M. Doi, J. Gunn, Z. Ivezic, S. Kent, R. Lupton, J. Munn, R. Nichol, J. Ostriker, D. Schlegel,
 M. Tegmark, D. Van den Berk, D. Weinberg and D. York,
``Cosmological parameter analysis including SDSS Ly-alpha forest and  galaxy bias: constraints 
on the primordial spectrum of fluctuations, neutrino mass, and dark energy'',
astro-ph/0407372.

\bibitem{smoot}
M.~Landriau and E.~P.~S.~Shellard,
``Large angle CMB fluctuations from cosmic strings with a comological constant,''
Phys.\ Rev.\ D {\bf 69} 023003 (2004),
astro-ph/0302166; \\
L.~Pogosian, M.~C.~Wyman and I.~Wasserman,
``Observational constraints on cosmic strings: Bayesian analysis 
in a three dimensional parameter space,''
astro-ph/0403268; \\
E.~Jeong and G.~F.~Smoot,
``Search for cosmic strings in CMB anisotropies,''
astro-ph/0406432.

\bibitem{Garcia-Bellido:2001ky}
J.~Garcia-Bellido, R.~Rabadan and F.~Zamora,
``Inflationary scenarios from branes at angles,''
JHEP {\bf 0201}, 036 (2002), hep-th/0112147.

\bibitem{Kaspi}
V.~M. Kaspi, J.~H. Taylor and M.~F. Ryba, Ap. J. {\bf 428}, 713 (1994); \\
S.~E. Thorsett and R.~J. Dewey Phys. Rev. {\bf D53}, 3468 (1996). 

\bibitem{Lommen}
A.N. Lommen, ``New Limits on Gravitational Radiation using Pulsars'',
astro-ph/0208572.

\bibitem{Alishahiha:2004eh}
M.~Alishahiha, E.~Silverstein and D.~Tong,
``DBI in the sky,''
Phys.\ Rev.\ D {\bf 70}, 123505 (2004), hep-th/0404084.

\bibitem{Giddings:2001yu}
S.~B.~Giddings, S.~Kachru and J.~Polchinski,
``Hierarchies from fluxes in string compactifications,''
Phys.\ Rev.\ D {\bf 66}, 106006 (2002), hep-th/0105097.

\bibitem{Klebanov:2000hb}
I.~R.~Klebanov and M.~J.~Strassler,
``Supergravity and a confining gauge theory: Duality cascades and
chiSB-resolution of naked singularities,''
JHEP {\bf 0008}, 052 (2000), hep-th/0007191.

\bibitem{DeWolfe:2004qx}
O.~DeWolfe, S.~Kachru and H.~Verlinde,
``The giant inflaton,''
JHEP {\bf 0405}, 017 (2004), hep-th/0403123; \\
S.~Kachru, J.~Pearson and H.~Verlinde,
``Brane/flux annihilation and the string dual of a non-supersymmetric  field theory,''
JHEP {\bf 0206}, 021 (2002), hep-th/0112197.

\bibitem{Silverstein:2003hf}
E.~Silverstein and D.~Tong,
``Scalar speed limits and cosmology: Acceleration from D-cceleration,''
Phys.\ Rev.\ D {\bf 70}, 103505 (2004), hep-th/0310221.


\end{thebibliography}
\end{document}